\documentclass[11pt,twoside]{article}
\usepackage{asp2010}

\resetcounters

\begin{document}

\title{Kinematic transitions and streams in galaxy halos}
\author{A.J.~Romanowsky,$^{1,2}$ J.A.~Arnold,$^2$ J.P.~Brodie$^2$, C.~Foster$^3$,
D.A.~Forbes$^4$,
H.~Lux$^5$,
D.~Mart\'inez-Delgado$^6$,
J.~Strader$^7$,
S.~Zibetti$^8$, and the SLUGGS team
\affil{$^1$Dept.\ Physics \& Astronomy, San Jos\'e State Univ., San Jose, CA 95192, USA}
\affil{$^2$Univ.\ of California Observatories, Santa Cruz, CA 95064, USA}
\affil{$^3$Australian Astronomical Observatory, North Ryde, NSW 1670, Australia}
\affil{$^4$Centre Astrophysics \& Supercomputing, Swinburne Univ., Australia}
\affil{$^5$School of Physics \& Astronomy, University of Nottingham, NG7 2RD, UK}
\affil{$^6$Astronomisches Rechen-Institut, Heidelberg, Germany}
\affil{$^7$Dept.\ Physics \& Astronomy, Michigan State Univ., East Lansing, MI, USA}
\affil{$^8$INAF - Osservatorio Astrofisico di Arcetri, I-50125 Firenze, Italy}
}

\begin{abstract}
The chemo-dynamics of galaxy halos beyond the Local Group may now be mapped out through
the use of globular clusters and planetary nebulae as bright tracer objects, along with
deep multi-slit spectroscopy of the integrated stellar light.
We present results from surveying nearby early-type galaxies, including evidence for kinematically
distinct halos that may reflect two-phase galaxy assembly.
We also demonstrate the utility of the tracer approach in measuring the kinematics of stellar substructures
around the Umbrella Galaxy, which allow us to reconstruct the progenitor properties and stream orbit.
\end{abstract}

\section{Stellar halos beyond the Local Group}

Galaxies may generically consist of two basic components:
an inner region that hosts in-situ star formation, and an outer accretion zone where satellite galaxies
fall in and disrupt into halo streams and substructures.
These components are reflected observationally by regions of distinct stellar densities,
kinematics, dynamics, and chemical abundances.
Thus these distinctions provide key information about the assembly histories
of galaxies, while dynamical analyses of coherent halo substructures  supply unique tests of the gravitational potential.

The Milky Way is a natural focal point for studying halo transitions, substructures, and satellites,
and for testing predictions from hierarchical galaxy formation theory
(e.g., \citealt{Carollo07,Law10,Kroupa10,McCarthy12}).
However, there are limitations to such comparisons owing to the expected stochasticity of galaxy assembly \citep{Cooper10}.
Indeed, the wealth of emerging information about the nearby galaxy M31
has revealed remarkable differences in the stellar halo properties of these two galaxies (e.g., \citealt{Font06b,Deason13}). 
This variation highlights  the need to extend detailed halo chemo-dynamical studies beyond one or two local spiral galaxies,
to galaxies of all types and environments, in order to obtain a robust understanding of structure formation in the universe.

Observing more distant stellar halos and substructures is very challenging, owing to the faintness of the starlight.
Progress has been made with imaging of stellar halos,
(e.g., \citealt{Mouhcine10,Martinez-Delgado12}), but direct spectroscopy remains more elusive, with success so far
only in the inner halos (out to tens of kpc; e.g., \citealt{Coccato10,Murphy11}).
For the outer halos, an effective solution was presaged by the discovery of the chaotic nature of the Milky Way halo \citep{Searle78}:
through the exploitation of bright tracer objects such as globular clusters (GCs) and planetary nebulae (PNe).
Current instrumentation permits the study of halo kinematics 
out to distances of $\sim$~50--100~Mpc \citep{Gerhard05,Misgeld11}.

GCs and PNe are not only mere kinematical tracers but also provide additional clues to galaxy formation by tracing
multiple underlying stellar populations. 
The PNe and metal-rich GCs trace the metal-rich field stars -- with the PNe possibly biased toward
intermediate-age populations -- while the metal-poor GCs trace the underlying classical metal-poor stellar halos
which are otherwise very difficult to observe directly (cf.\ \citealt{Park13}).
The potential of this method is demonstrated by the
kinematical differences detected among these subpopulations within single galaxies
\citep{Coccato13}, and by expectations for galaxy merger remnants to host dynamical subcomponents that are 
traceable to distinct parts of their progenitors \citep{Hoffman10}.

\section{Early-type galaxies}

Some giant early-type galaxies (ellipticals and lenticulars)  host kinematically 
decoupled or distinct cores (KDCs), with strong rotational transitions on scales of $\sim$~0.5~kpc
\citep{Krajnovic11}. These features may be residues of central starbursts,
and are relatively rare, while one could expect transitions to be more common 
at larger radii, reflecting generic two-phase assembly histories.
The accreted component is predicted to be substantial outside of a few kpc, and to
increase in importance with galaxy mass \citep{Oser12,Lackner12}.

Halo kinematic transitions in early-types are now being explored through two major surveys:
one using the PN.Spectrograph \citep{Coccato09},
and the other called SLUGGS (Brodie et al., in preparation).
The latter is based on GCs \citep{Pota13a}, along with a novel multi-slit technique for
mapping out integrated stellar spectroscopy in two dimensions to
$\sim$~3 effective radii ($R_{\rm e}$; \citealt{Proctor09}).

\begin{figure}[!ht]
\plotone{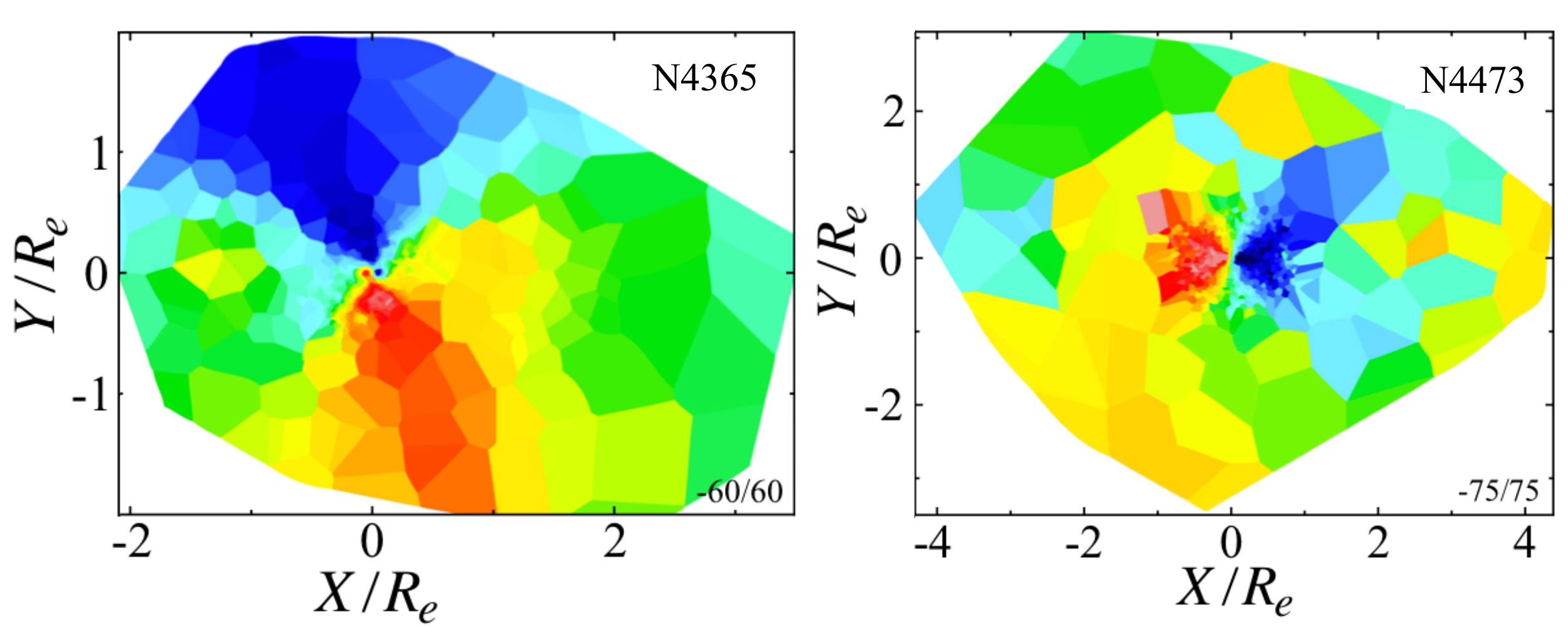}
\caption{Smoothed wide-field stellar velocity maps of two nearby early-type galaxies from the SLUGGS survey \citep{Arnold14}.
The $X$ axes are aligned with the photometric major axes.
NGC~4365 has a well-known KDC embedded in a ``rolling'' minor-axis rotation pattern, 
which is now seen to extend to large radii.  This galaxy may be a major-merger remnant, with the
rolling rotation tracing a faded polar ring that originated in progenitor disk spin \citep{Hoffman10}.
NGC~4473 has an abrupt kinematical change at $\sim 1 R_{\rm e}$ to a kinematically distinct halo, which shows
simultaneous major and minor axis rotation (see also \citealt{Foster13}).
}\label{fig1}
\end{figure}

These surveys have revealed a remarkable variety of rotation profiles and kinematic twists in the galaxy
halos that deviate from the patterns found in their inner regions.  Some cases
resemble giant KDCs, and may be considered kinematically distinct halos (KDHs;
\citealt{Foster13}; Figure~\ref{fig1}).
In a similar vein, H{\sc i} disks and rings around early-type galaxies appear to be frequently decoupled from their
central regions \citep{Serra14},
while large-radius transitions in density and stellar populations have also been found
(e.g., \citealt{Forbes11,Huang13,Petty13}) -- all of which may support the two-phase paradigm.

Interpreting the large-radius kinematical  data  will require detailed dynamical modeling, 
and comparisons to simulations of galaxy formation.
Such simulations in a cosmological context have shown kinematic
diversity in their outer parts \citep{Wu14}, with further work needed to
connect these variations to specific assembly histories.

\section{Stellar streams}

While kinematic transitions in galaxy halos may reflect the jumbled debris from 
multiple accretion events spanning a Hubble time, 
the cases of visible stellar streams are valuable in order to study the dynamics of accretion.
Such streams may be more common than the canonical massive, gas-rich polar rings
owing to the lower satellite masses and to the morphology--density relation which
implies that ongoing accretion events will generally be gas poor.
This presents again the observational obstacle of low surface brightness,
with bright tracers as the escape clause, 
along with a new technique for using blends to extend individual-star spectroscopy
out to $\sim$~5~Mpc \citep{Theakanath14}.
The tracers approach with PNe was demonstrated for the Giant Southern Stream in M31
\citep{Merrett03}, and has been applied with GCs to streams at $\sim$~20~Mpc distances
\citep{Romanowsky12a,Blom14}.

\begin{figure}[~ht]
\plotone{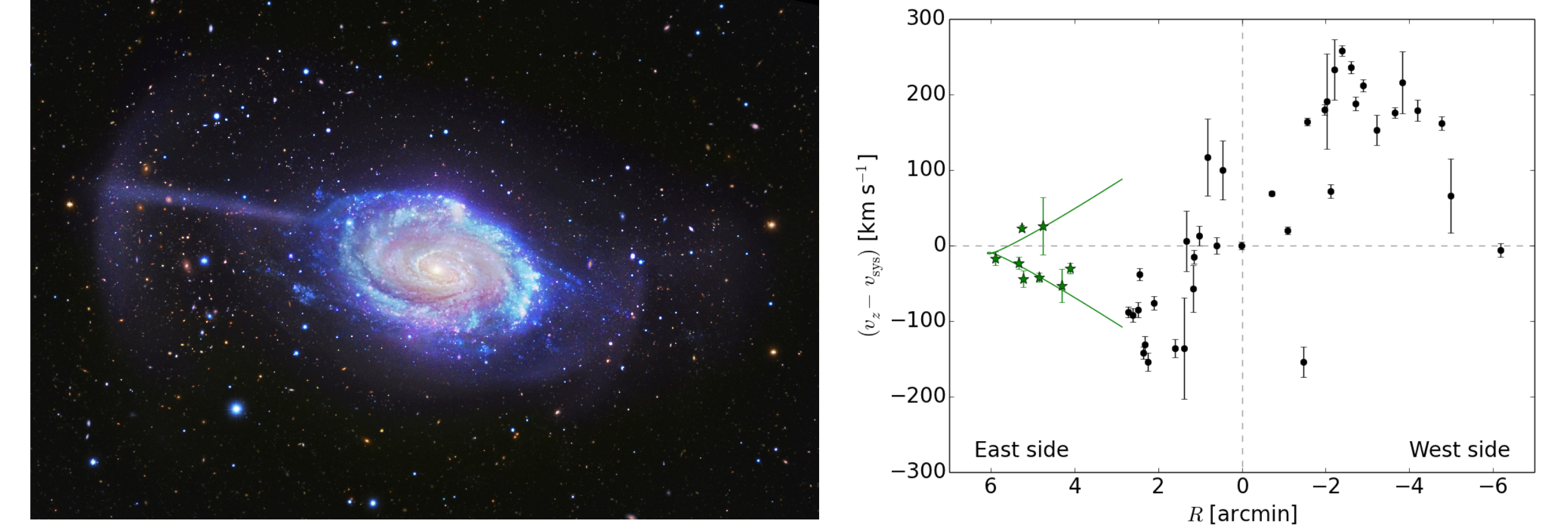}
%\plottwo{gabany_v4_high_contrast_n4651_cropped-rotated.jpg}{umbrella_RV2.eps}
%\plotfiddle{umbrella_RV2.eps}{10cm}{0}{20}{20}{0}{0}
\caption{The Umbrella Galaxy, NGC~4651.
{\it Left:} A color image combining Subaru/Suprime-Cam with amateur telescope data.
A narrow stream terminates in an arc to the left, with several plumes and shells to the right of the galaxy.
These features, in combination with kinematics measured from GC and PN tracers, have allowed a reconstruction
of the stream progenitor and orbit. 
{\it Right:} Position--velocity phase-space of tracers, where green star symbols are associated with the umbrella feature.
The curves show a model of the shell caustics \citep{Sanderson13}.
}\label{fig2}
\end{figure}

For maximal stream constraints, a combination of GCs and PNe should be used,
as recently done in the Umbrella Galaxy, NGC~4651
(Foster et al., in preparation; Figure~\ref{fig2}).
Here a handful of velocities in the shell-and-stream structure were sufficient to delineate its
trajectory in phase-space and to enable a fit with a simple orbit model.
The substructure photometry and dynamics indicate 
a disrupted gas-poor nucleated dwarf galaxy on a very eccentric, fairly polar orbit, analogous to
the Giant Southern and Sgr streams in the Local Group.
More detailed modeling of the host galaxy disk perturbation (cf.\ \citealt{Purcell11}), and of the stream dynamics,
could provide constraints on the dark matter distributions of both the satellite and its host.

\bibliography{romanowsky}

\end{document}